\documentclass[]{jpconf}

\usepackage{graphicx}
\usepackage{textcomp}
\usepackage{comment}
\usepackage[english]{babel}
\usepackage{miller}

\usepackage{caption}
\usepackage{subfig}
\usepackage{amsmath}
\usepackage{tikz}

\usepackage[sort&compress,numbers]{natbib} % also for aps-rev bibstyle
\def\newblock{\hskip .11em plus .33em minus .07em} %  required by natbib in combination with certain bibliography styles
\usepackage[unicode=true, 
 bookmarks=true,bookmarksnumbered=false,bookmarksopen=false,
 breaklinks=true,pdfborder={0 0 1},backref=false,colorlinks=true]
 {hyperref}
\hypersetup{pdftitle={Physics-based Simulation Models for EBSD}, pdfauthor={Aimo Winkelmann}}

\begin{document}
\title{Physics-based Simulation Models for EBSD: \\ 
Advances and Challenges}

\author{Aimo Winkelmann}
\address{Bruker Nano GmbH, Am Studio 2D, 12489 Berlin, Germany}
\ead{aimo.winkelmann@bruker.com}

\author{Gert Nolze}
\address{BAM, Federal Institute for Materials Research and Testing, Unter den Eichen 87, 12205 Berlin, Germany}

\author{Maarten Vos}
\address{Atomic and Molecular Physics Laboratories, Research School of Physics and
Engineering, The Australian National University, Canberra, Australia 0200}

\author{Francesc Salvat-Pujol, Wolfgang Werner}
\address{Vienna University of Technology, Wiedner Hauptstrasse 8-10, A-1040 Vienna, Austria}

\begin{abstract}
EBSD has evolved into an effective tool for microstructure investigations in the scanning electron microscope. 
The purpose of this contribution is to give an overview of various simulation approaches for EBSD Kikuchi patterns and to discuss some of the underlying physical mechanisms.
\end{abstract}

\section{Introduction}

Access to suitable microscopic characterization methods of crystalline phases is a necessary prerequisite in order to develop new applications of advanced materials. In this context, diffraction methods are essential tools.
The method of electron backscatter diffraction (EBSD) is based on the measurement of Kikuchi patterns \cite{alam1954prsl} 
which appear in the angular distribution of backscattered electrons (BSE) in the scanning electron microscope (SEM). 
In the context of the EBSD method, these patterns are also called electron backscatter diffraction patterns (EBSP), or backscattered Kikuchi patterns (BKP) with the extension to transmission Kikuchi diffraction (TKD) in the SEM.
Over the last two decades, EBSD has evolved into an effective tool for microstructure investigations
\cite{venables1973pm,schwartzEBSD2,wilkinson1997m,schwarzer1997micron,dingley2004jm,randle2008ae} 
with a wide field of applications in orientation mapping \cite{adams1993mt}, grain- and phase-boundary characterization, local phase identification, and lattice strain determination \cite{wilkinson2006um,maurice2008jm,wilkinson2012mt}.
Because a range of physical and chemical properties of modern materials is directly affected by the crystal structure, there is an evident need for characterization techniques like EBSD which are locally sensitive to crystal symmetries on the relevant length-scales of the material's microstructure, possibly down to the nanoscale \cite{small2002jm,keller2012jm,trimby2012um,brodusch2013jm}. 

\begin{figure}
\label{fig:principle}
\captionsetup{type=figure}% tell subfig package that this is a figure
\centering

    \subfloat[Schematic of an EBSD setup]{
    \label{fig:setup}
	\includegraphics[width=5.76cm]{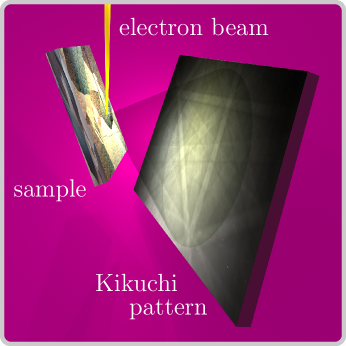}
    } 
    \subfloat[Kikuchi pattern of GaP with the main zone axes and lattice planes indicated]{
    \label{fig:GaP_exp}
    \includegraphics[width=8.0cm]{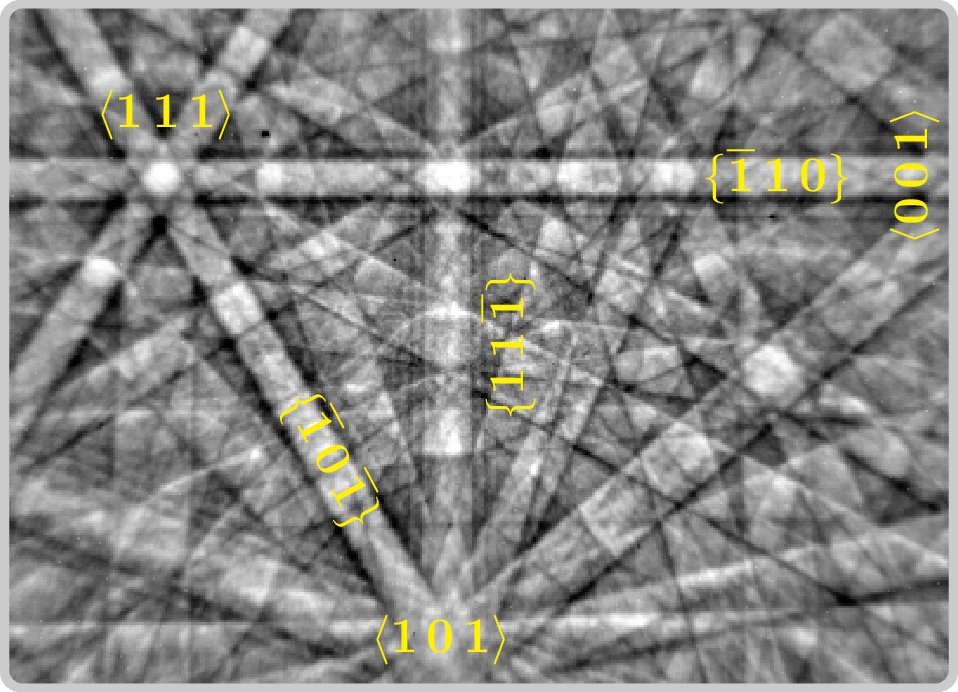}
    } 

\caption{Principle of Kikuchi pattern measurement in the SEM. The primary electron beam is focused on a tilted  sample, and the Kikuchi pattern is measured in gnomonic projection on a planar detector. 
The combined local resolution of the electron beam and the interaction volume for Kikuchi pattern formation determines the resolution that is available for spatially resolved orientation mapping.}
\end{figure}

Experimentally, EBSD is conceptually simple: in principle only a phosphor screen imaged by a sensitive CCD camera is needed, see Fig. \ref{fig:setup}. 
The geometry of the Kikuchi patterns is governed by the gnomonic projection of the Bragg reflection conditions of a point source inside a crystal \cite{kossel1936annphys,gajdardziska1991aca}.
However, this does not give any information on the details of the observed intensity distribution near a Kikuchi band, and the explanation of the exact physical Kikuchi pattern formation process is considerably more complicated.
Kikuchi pattern formation is based on the concept of atomic emitters of electron waves which are incoherent with respect to the incident beam \cite{reimersem} and the multiple scattering of these waves inside a crystal gives rise to the characteristic Kikuchi pattern intensity distribution. 
Because of the strong effects of multiple scattering and absorption, the quantitative calculation of the backscattered diffraction pattern has to be based on the dynamical theory of electron diffraction.

Due the increased information content which is provided by dynamical electron diffraction in EBSD Kikuchi patterns, the detailed physics-based simulation and analysis of the respective effects is becoming more and more important \cite{winkelmann2007um,winkelmann2009ebsd2}. 
For example, Kikuchi patterns are intrinsically sensitive to the point-group symmetry of a crystal \cite{marthinsen1988aca,babakishi1991jac}.
With the huge increase of the commonly available computing power, it has become possible to routinely carry out quantitative simulations of high-resolution Kikuchi patterns with an efficiency that now makes it possible to directly compare complete experimental Kikuchi patterns with theoretical simulations.
Typical applications of dynamical diffraction simulations include EBSD-based lattice strain analysis \cite{villert2009jm,wilkinson2012mt}, EBSD image segmentation and orientation determination \cite{park2013icip,callahan2013mm,chen2015arx}, and the analysis of orientation relationships in phase transformations \cite{nolze2014crt}.
 With respect to non-centrosymmetric phases, point-group sensitive orientation mapping has also been demonstrated \cite{winkelmann2015apl}.
The range of possible applications of dynamical EBSD pattern simulations includes even space-group sensitivity, which has been demonstrated by discrimination of left-handed and right-handed quartz using EBSD Kikuchi patterns \cite{winkelmann2015um}.

SEM based diffraction can also be observed via Electron Channeling Patterns (ECP) \cite{joy1982jap} which measure Kikuchi patterns of backscattered electrons resulting from the diffraction of the incident beam. Due to its close relationship to EBSD, the results presented here also largely apply to ECP, which is a particularly useful method for direct imaging of dislocations in the SEM \cite{wilkinson1997m}. This type of application which has recently received renewed interest \cite{nareshkumar2012prl,zaefferer2014am} with decisive advances in the necessary quantitative image simulations \cite{picard2014um}.

The purpose of this contribution is to give an overview of simulation approaches for Kikuchi patterns and some of the underlying physical mechanisms. In this way, the present paper aims to show that EBSD (and also ECP) in combination with Kikuchi pattern simulations provides an important tool for advanced analysis of crystalline phases in the SEM.

\section{Models for Kikuchi pattern simulation}
\label{sec:models}

In this section we review the existing models for EBSD-related Kikuchi pattern simulation in order to critically point out their specific advantages and limitations.

The calculation of all the simulated Kikuchi patterns was carried out using the software \mbox{\emph{ESPRIT DynamicS}} (Bruker Nano) which implements the Bloch wave approach for EBSD pattern simulation \cite{winkelmann2007um,winkelmann2009ebsd2}.
In detailed Kikuchi pattern simulations, a typical number of about $1000$ or more reciprocal lattice vectors $hkl$ is taken into account. These are selected according to the reciprocal lattice vector length $d^*_{hkl}=1/d_{hkl}$ (typically $d^*_{hkl} < 1/0.05 \ldots 1/0.035$ nm$^{-1}$) and the relative strength with respect to the  largest structure factor amplitude $|F|_{\mathrm{max}}$ (typically 15\%...10\% or less). 
The simulations presented here neglect experimental excess-deficiency effects which are related to the incident beam direction \cite{winkelmann2008um}. 
To quantify the agreement between two Kikuchi patterns, we use the normalized cross-correlation coefficient $r$ \cite{gonzalez2002,tao2005mm}: 

\begin{equation}{\label{eqn:r}}
r = \frac{  \sum\limits_{i,j} \left[ f(i,j)-\bar{f} \, \right] \cdot \left[ w(i,j)-\bar{w}\,\right] } { \sqrt{ \sum\limits_{i,j} \left[ f(i,j)-\bar{f} \, \right] ^2} \, \cdot \,  \sqrt{ \sum\limits_{i,j} \left[ w(i,j)-\bar{w} \, \right] ^2} }
\end{equation}
where $f(i,j)$ and $w(i,j)$ are the pixel intensity values of the corresponding regions of interest (ROI) in the two patterns to be compared, while $\bar{f}$ and $\bar{w}$ are the mean values in these ROIs. 
The absolute value of $r$ is in the range between 0 and 1 and does not depend on scale changes in the intensity of both patterns. 
From our experience, values of $r>0.6...0.7$ indicate visually convincing fits between experimental and simulated values. 
For pattern matching or model discrimination via $r$, the statistically necessary levels of $r$ could be lower or higher than these values, depending on the details of the application.

In the following sections we proceed to evaluate the capability of different theoretical models to describe the experimentally observed Kikuchi patterns.

\subsection{The geometrical model}

\begin{figure}   %
\centering
\includegraphics[width=7.0cm]{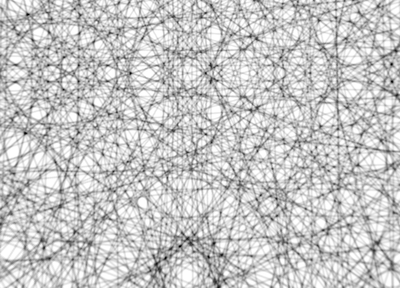}
\caption{Gnomonic projection of the possible Bragg conditions in GaP for $d_{hkl}>0.5$ \AA\, and $|F|_{hkl}>0.15 |F|_{max}$ for the same experimental conditions as in Fig.\,\ref{fig:GaP_exp}.}
\label{fig:bragg}
\end{figure}

In the simplest case, the Kikuchi patterns are interpreted with the help of the gnomonic projection of Kossel cones \cite{young1972jap,kossel1936annphys} which are formed by all possible directions of Bragg reflection from a lattice plane with normals \hkl(hkl) and \hkl(-h-k-l). This does not include any information about band intensities.
In this model, the width of a Kikuchi band is given by twice the Bragg angle $\theta_B$:
\begin{equation}\label{Bragg1}
\sin \theta_B = \frac{n\cdot\lambda}{2\cdot d_{\hkl(hkl)}}.
\end{equation}
In the case of EBSD the resulting Bragg angle is in the order of only a few degrees because of the relative sizes of the electron wavelength $\lambda$ and the unit cell dimensions. 
Diffraction orders with $n>1$ in (\ref{Bragg1}) describe the higher-order interferences that form additional parallel lines outside the main band. 
A geometrical simulation for GaP displaying the Kossel cone intersections of all possible Bragg reflections with $d_{hkl}>0.05$nm and $|F|_{hkl}>0.15 |F|_{max}$ can be seen in Fig. \ref{fig:bragg} (the same beam selection criteria will be applied in all the following simulations shown here).
While the actual agreement with the experiment is very limited, the entangled network of Kossel cone projections gives a peculiarly indirect view of the structure of reciprocal space and its effects on EBSD Kikuchi patterns.
The most notable features are the characteristic circular areas surrounding some important zone axes like \hkl<111>, which are related to the "Higher Order Laue Zone" (HOLZ) rings seen also in the experimental GaP pattern \cite{michael2000um}. 
It is useful to note that all the reflections shown in Fig. \ref{fig:bragg} will also enter into the dynamical simulation discussed later, where the influence of each possible reflection is considered quantitatively, in contrast to the purely geometrical drawing of Kossel-cone hyperbolas.
The simple Bragg-reflection interpretation also neglects any non-centrosymmetric effects since the Bragg angle is the same for $d_{\hkl(hkl)}$ and $d_{\hkl(-h-k-l)}$ and the respective Kossel cones are mirror-symmetric with respect to the relevant lattice plane.

\subsection{The kinematical model}

\begin{figure}
\captionsetup{type=figure}% tell subfig package that this is a figure
\centering

    \subfloat[Intensity at Bragg reflection scaled by $|F_{hkl} |^2 $ ]{
        %\label{subfig:correct}
        \includegraphics[width=7.0cm]{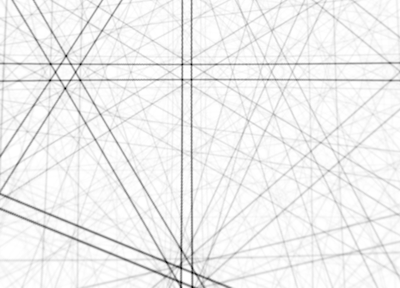}
    } 
    \hspace{0.2cm}
    \subfloat[Intensity at Bragg reflection scaled by $|F_{hkl} |$ ]{
        %\label{subfig:notwhitelight}
        \includegraphics[width=7.0cm]{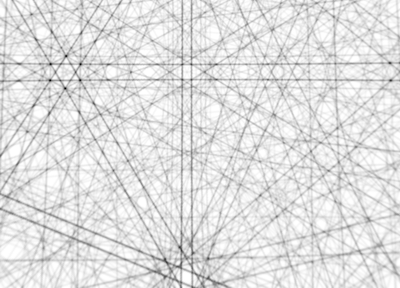}
    } 

\caption{Gnomonic projection of the Bragg condition in GaP. The darkness level of the lines at the Bragg conditions is scaled differently as a function of the structure factor amplitude.}
\label{fig:bragg_kin}
\end{figure}

As can be seen by comparison of Fig.\ref{fig:GaP_exp} and Fig.\ref{fig:bragg}, not all of the geometrically possible bands are equally visible in the experimentally measured Kikuchi pattern. 
In order to account for this, the kinematical intensity $I_{hkl}$ of a Bragg reflection $hkl$ is often used empirically to describe the different relative intensities of Kikuchi bands. As is well known from X-ray diffraction, $I_{hkl}$ is given by the square of the absolute value of the structure factor $|F|_{hkl}^2$ in the absence of $hkl$-dependent absorption effects:
\begin{equation}\label{eqn:I}
I_{hkl}\propto F_{hkl}^*F_{hkl}
\end{equation}
A corresponding scaling of the Bragg reflection lines is shown in Fig. \ref{fig:bragg_kin}, which allows to roughly estimate the relative intensities of the experimentally observed Kikuchi bands in Fig.\,\ref{fig:GaP_exp}. 

Assignment of a kinematic intensity $I_{hkl}$ and $I_{\bar{h}\bar{k}\bar{l}}$ to the two related Bragg reflections of a Kikuchi band is likewise still neglecting non-centrosymmetric effects. In the kinematic model, there is no possibility to differentiate between $I_{hkl}$ and $I_{\bar{h}\bar{k}\bar{l}}$ due to Friedel's rule, which states that both these intensities are equal in the absence of anomalous absorption.
Thus, the kinematical EBSD model is sensitive to Laue-symmetry only. 

\begin{figure}
\captionsetup{type=figure}% tell subfig package that this is a figure
\centering

    \subfloat[Band intensity scaled by $|F_{hkl} |^2 $ , $r_{ROI}=0.507$ ]{
        %\label{subfig:correct}
        \includegraphics[width=7.0cm]{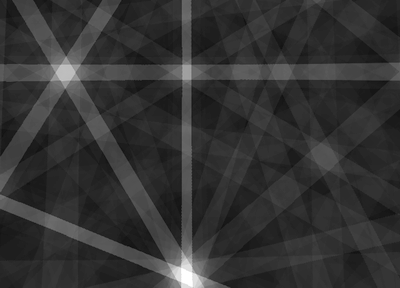}
    } 
    \hspace{0.2cm}
    \subfloat[Band intensity scaled by $|F_{hkl} | $ , $r_{ROI}=0.580$ ]{
        %\label{subfig:notwhitelight}
        \includegraphics[width=7.0cm]{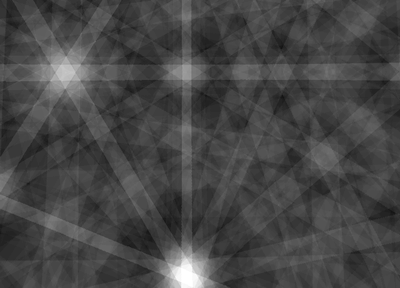}
    } 

\caption{Gnomonic projection of GaP Kikuchi bands with step profiles $B_K$ according to equation \ref{eqn:boxprofile}. The band intensity is assumed to be (a)  $ \propto |F_{hkl}|^2$, or (b) $\propto |F_{hkl}|$. Correlation coefficients $r$ are taken with respect to the experimental pattern in Fig.\ref{fig:GaP_exp}.}
\label{fig:box}
\end{figure}

The angle-dependent Kikuchi band profile intensity can be described by a function $B(\vartheta)$, where $\vartheta$ is measured from the relevant lattice plane. 
The general form of $B(\vartheta)$ is determined by the distance of $\vartheta$ to the Bragg angle.
For large angles, we observe the average background intensity, since we are away from diffraction effects which are strongest near the Bragg angle. 
In the region near the Bragg angle, the Kikuchi band intensity is modulated and can go below or above the background for different angles.
To approximate the complicated experimental Kikuchi band profile, a constant intensity $I$ can be empirically assigned to the full geometrical band width, resulting in bands with step profiles $B_K(\vartheta)$:
\begin{equation}\label{eqn:boxprofile}
B_K(\vartheta) =
\begin{cases}
    I,				& \text{if } |\vartheta| \leq \vartheta_B\\
    0,              & \text{otherwise}
\end{cases}
\end{equation}
This formula introduces an additional experimental observation, namely that the Kikuchi bands in EBSD usually (but not always) show a higher intensity than the background. From within the kinematical model itself, however, the actual intensity distribution $B(\vartheta)$ of a Kikuchi band cannot be predicted at all. This also includes the constant $I$ for the band intensity. The obvious choice seems to be the kinematical intensity $I_{hkl}$. 
In order to check the visual agreement of this assumption with the experiment, in Fig.\,\ref{fig:box} we have displayed the Kikuchi bands $B_K$ with grey values in one case proportional to the kinematical intensity  $I_{hkl}\propto |F|^2$, and in the other case proportional to the structure factor amplitude $|F|$. We found correlation coefficients of $r(|F|^2)=0.51$ and $r(|F|)=0.58$ for both these types of simulations showing that the linear dependence is the better approach and that assignment of $I_{hkl}$ is not optimal. (The correlation coefficients have been calculated for a ROI covering the right side of the patterns, thereby excluding the high intensity zone axes on the left side which otherwise would distort the comparison.) 

We would like to stress that the kinematical EBSD models are limited in several important aspects: 
\begin{enumerate}
\item The kinematical intensity $I_{hkl}$ actually describes the intensity of an incident plane wave beam which comes from infinity, is coherently Bragg reflected by single scattering processes at the crystal atoms and goes again to infinity to be detected in a typical spot-like diffraction pattern. This is a very different experimental situation than the case of an internal point source inside the crystal which emits an electron wave that is multiply scattered and then detected at infinity outside the crystal. Extensions of results from one situation to the other are not necessarily valid. 
\item The kinematical model cannot provide an actual formula for the Kikuchi band profile. Instead, the kinematical intensity $I_{hkl}$ is assigned purely empirically. 
While this simple model might usefully approximate a typical standard situation in EBSD and gives a rough impression about the intensity distribution within an Kikuchi pattern, it totally fails to describe effects like Kikuchi band contrast inversion where one actually observes dark Kikuchi bands  \cite{alam1954prsl,winkelmann2010um}.
\item The pairs of Bragg reflections $hkl$ and $\bar{h}\bar{k}\bar{l}$, which are relevant for a Kikuchi band, independently contribute to the final pattern in a simple sum of all "step band" intensities $B_K$. This results in unrealistic intensity maxima especially along low-indexed zone axes where many bands are crossing each other. 
\item The kinematical intensity scales with the square of the structure amplitude  $I_{hkl}\propto|F_{hkl}|^2$. When compared to experimental Kikuchi patterns, however, a linear dependence on the structure factor amplitude actually seems to provide a better fit, as we have seen above. 
The fact that the kinematical intensity $I_{hkl}$ does not give the best agreement of the simulations using $B_K$-profiles exposes the severe limitation of the kinematical theory in the description of the Kikuchi band profile.
Depending on the choice of $I$, the modified intensity decay from the stronger to the weaker reflections  causes a different relative importance of specific simulated pattern features. 
This could influence the results when the $B_K$-model is applied in correlation-based strain determination, where various dependencies on $F_{hkl}$ can be selectively chosen \cite{abouras2008crt,kacher2009um,britton2010um,alkorta2013um}.
In view of the general deficiencies of the kinematical model, we have not looked for other functional dependencies that might provide an even better empirical fit to the experiment.
\end{enumerate}
\subsection{The two-beam dynamical profile}

While the kinematical EBSD model neglects most of the physics which is involved in the detailed Kikuchi band formation, the dynamical theory of electron diffraction can provide a detailed description of experimentally measured Kikuchi patterns.
A simplified dynamical theory of the Kikuchi pattern profile is given in \cite{reimersem}, where an analytical band shape is derived based on the two-beam approximation.
Assuming only single-element backscatterers placed on Bravais lattice positions, this band shape expresses the variation of the backscattered electron intensity near a Bragg reflection with reciprocal lattice vector $\mathbf{g}$ qualitatively as:
\begin{equation}
\label{eq:reimereta}
B_A(\vartheta) \propto -\frac{w_{\mathbf{g}}+a_{\mathbf{g}}}{1+w^2_{\mathbf{g}}-a^2_{\mathbf{g}}} + \ldots
\end{equation}
where $w_{\mathbf{g}}=w_{\mathbf{g}}(\vartheta, F_{hkl},\lambda) $ is a normalized deviation parameter with $w_{\mathbf{g}}=0$ at the Bragg condition,  and $a_{\mathbf{g}}$ contains absorption parameters for the reflection $\mathbf{g}$. 
In the dynamical model, absorption parameters turn out be important for the actual shape of the Kikuchi band profile and they can even contribute to an inverted band shape. This illustrates another deficiency of the kinematical model, which does not consider these absorption effects at all. For more details on the simplified two-beam profile, we refer the reader to Chapter 9 in \cite{reimersem}. 
\begin{figure}   %
\captionsetup{type=figure}% tell subfig package that this is a figure
\centering

    \subfloat[Experiment]{
        \label{fig:GaPExpDyn}
        \includegraphics[width=7.0cm]{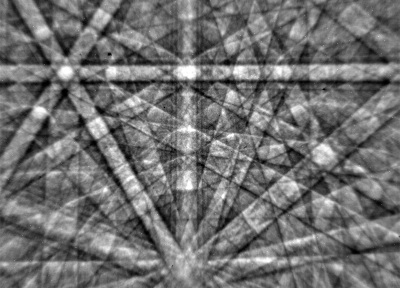}
    } 
    \hspace{0.2cm}
    \subfloat[Sum of two-beam dynamical profiles, $r=0.71$]{
        \label{fig:tba}
        \includegraphics[width=7.0cm]{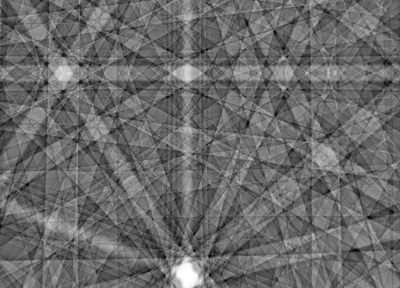}
    } 

    \subfloat[Many-beam dynamical simulation, best fit for point group $\bar{4}3m$ , $r=0.81$]{
        \label{fig:GaP081}
        \includegraphics[width=7.0cm]{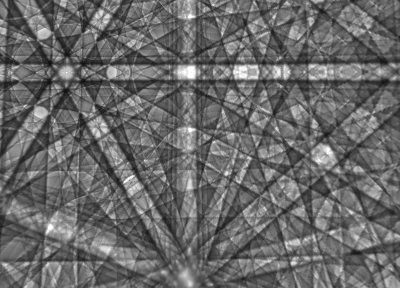}
    } 
    \hspace{0.2cm}
    \subfloat[Many-beam dynamical simulation, \mbox{$r=0.78$}, Laue-group $m\bar{3}m$ equivalent orientation \mbox{(90$^\circ$ rotation around $\bar{4}$-axis)}]{
        \label{fig:GaP078}
        \includegraphics[width=7.0cm]{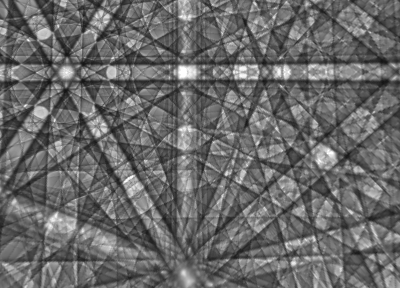}
    } 
\caption{Comparison of dynamical simulations to an experimental GaP pattern.}
\label{fig:expdyn}
\end{figure}

In Fig. \ref{fig:tba}, the Kikuchi pattern of GaP is simulated using a sum of analytical two-beam profiles $B_A$ according to equation \ref{eq:reimereta}. 
While the computational effort of this approach is still rather low, the resulting Kikuchi pattern already shows a much better agreement with the experiment than any of the kinematical approximations discussed above.
A main deficiency of this approach is again seen in the zone axes, where the fine structure is not reproduced faithfully due to the simple sum of band intensities with no interference. Also, in the theoretical derivation of this model, the actual distribution of the different chemical elements in the unit cell is not taken into account exactly. 
In view of the limits of the kinematical models discussed above, the analytical two-beam profile is a better standard choice for simplified pattern simulations in the absence of the full many-beam dynamical calculations which are discussed next.  

\subsection{Many-beam dynamical pattern simulation}

Realistic intensity simulations of EBSD patterns have been introduced using the Bloch wave approach of the dynamical theory of electron diffraction \cite{winkelmann2007um,winkelmann2009ebsd2,villert2009jm,callahan2013mm}. 
This approach calculates the probability that an electron will leave the crystal in a specific direction after having been incoherently scattered from the incident beam at an atomic position inside the crystal and then having undergone multiple coherent scattering in the crystal in the presence of absorption effects.

In order to compare the typical quality of dynamical EBSD simulations using the many-beam approach, in Fig.\,\ref{fig:GaP081} the resulting best fit intensity distribution is displayed in comparison to the experimental pattern in Fig.\,\ref{fig:GaPExpDyn}. 
As can be seen, the numerous specific features around zone axes and within bands are well reproduced by the many-beam simulation.

In the middle of the pattern, \hkl{111} becomes visible as a vertical band. 
The intensity of this band is asymmetrically shifted to the right, as can be even seen by eye on close inspection. 
This specific asymmetry is an effect of the crystal polarity in the non-centrosymmetric zinc-blende type structure of GaP, which can also be seen quantitatively in the cross-correlation coefficients:
The best-fit simulation gives a normalized  $r=0.81$. 
The Laue-group equivalent orientation in Fig.\,\ref{fig:GaP078} has an additional 90\textdegree{} rotation around the $c$-axis and gives a worse value of $r=0.78$, with the difference of $\Delta r=0.03$ being statistically significant. 

\begin{figure}   %
\centering
\includegraphics[width=10cm]{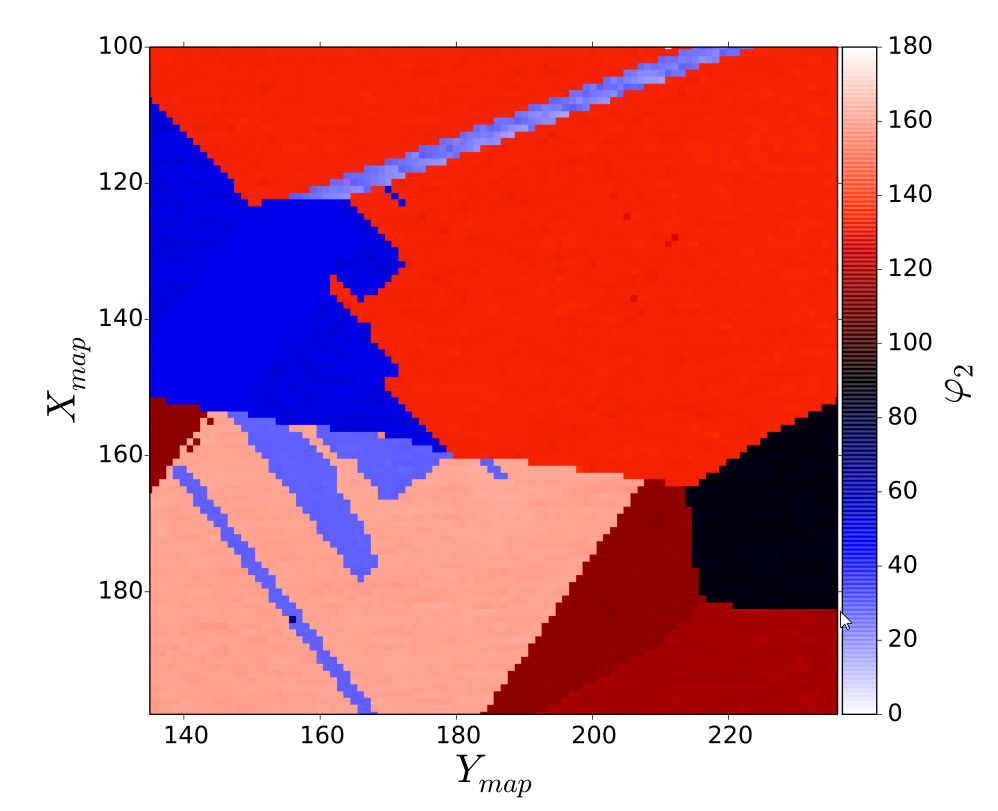}
\caption{Point-group sensitive orientation map for polycrystalline GaP \cite{winkelmann2015apl}, showing the value of the Euler angle $\varphi_2$. In the GaP point group $\bar{4}3m$, $\varphi_2$ can assume values from 0 to 180\textdegree{}, whereas the corresponding Laue group $m\bar{3}m$ limits $\varphi_2$ to the range 0..90\textdegree{}}
\label{fig:mapping}
\end{figure}
The different values of $r$ with respect to the two Laue-group equivalent orientations clearly demonstrates the sensitivity of the dynamical simulations to the actual point group of the crystal in which these two orientations are not equivalent. 
Such a point-group sensitive orientation analysis is beyond the limits of the conventional procedure that is applied up to now in EBSD systems. 
As a unique new application of dynamical diffraction simulations, in Fig.\ref{fig:mapping} we show results of absolute orientation mapping for polycrystalline GaP \cite{winkelmann2015apl}. 
The conventional Laue-group-specific analysis of EBSD Kikuchi patterns has to leave the absolute orientation of the GaP crystal undetermined by 90$^\circ$ around \hkl<001>. This indeterminacy is resolved by the point-group sensitive pattern matching approach.

Summarizing this section, the kinematic EBSD interpretation is limited to a description of the Kossel-cone geometry of Bragg angles and to a rough estimation of the ordering of reflections according to their kinematical intensity. Still, this is useful for the selection of the strongest reflectors $hkl$ for indexing of EBSD patterns. The kinematical models for EBSD are completely insensitive to the absence of an inversion center in the crystal structure.
To realistically describe EBSD Kikuchi patterns, one needs to use models which apply the dynamical theory of electron diffraction. The lack of an inversion center in the crystal structure is manifested in non-symmetric Kikuchi band profiles for polar lattice planes.

\section{Physical mechanisms}

The dynamical diffraction simulations presented above are based on certain assumptions about the physical pattern formation mechanisms. In this section, some of these fundamental effects are presented.

\subsection{Transfer of momentum in crystal scattering}

\begin{figure}   %
\centering
\includegraphics[width=11cm]{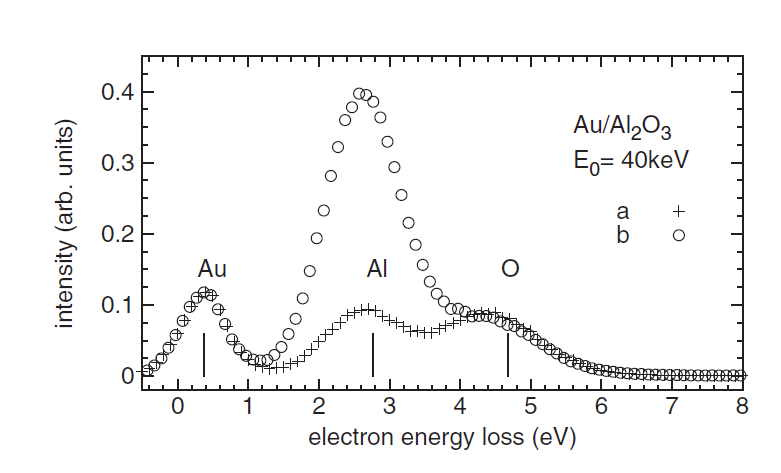}
\caption{Measured energy spectra of backscattered electrons from Au/Al$_2$O$_3$ for two different exit directions \cite{winkelmann2011prl}. 
The vertical lines indicate the energy transfer that would be expected from a classical elastic "billiard ball" collision between a single electron of 40keV and the respective atom.}
\label{fig:recoil}
\end{figure}

The electrons which are observed on the EBSD screen have undergone a substantial change in their direction of movement. This implies a corresponding transfer of momentum $\mathbf{q}$. Coherent elastic scattering from the crystal describes the situation when the electron does not excite any internal degrees of freedom of the crystal and a macroscopic mass $M_C$ of the crystal takes up the momentum. This results in an infinitesimal value of the recoil energy $E_R^C=|q|^2/2M_C$. However, there is also the possibility that a single atomic mass $M_A$ takes up the momentum, which then results in a measurable recoil energy even for electron scattering at typical SEM energies \cite{went2008nimb}. 
The probability of coherent elastic scattering is strongly suppressed if the respective single-atom recoil energy $E_R^A=|q|^2/2M_A$ exceeds typical phonon energies \cite{lipkin1973qm}. This is the case for a typical EBSD geometry, where coherent backscattering from the crystal can thus be usually completely neglected and spot diffraction patterns of the incident beam are absent. Exceptions can be observed by RHEED (reflection high energy electron diffraction) patterns in EBSD at glancing angles or in transmission Kikuchi diffraction (TKD) near the incident beam direction, when spot patterns can be observed that result from coherent elastic scattering. 
Due to the dependence of the recoil energy on the atomic mass, electron spectroscopy can be applied to differentiate between different scattering atoms. This is shown in Fig.\,\ref{fig:recoil}, where Au atoms on a sapphire ($Al_2O_3$) surface have been measured for two exit angles \cite{winkelmann2011prl}. The three peaks corresponding to Au, Al, and O can be clearly distinguished at their expected positions. The two different curves in Fig.\,\ref{fig:recoil} show the influence of diffraction effects due to different measurement geometries. 
In order to prove that Kikuchi patterns result from these incoherent recoil scattering events, we measured element-resolved Kikuchi band profiles and compared them with dynamical simulations \cite{winkelmann2011prl}. In principle, the recoil loss in the electron-atom collision makes it possible to measure separate Kikuchi patterns for different chemical elements in a compound phase.

\subsection{Energy spectrum of backscattered electrons}
\begin{figure}   %
\centering
\includegraphics[width=11cm]{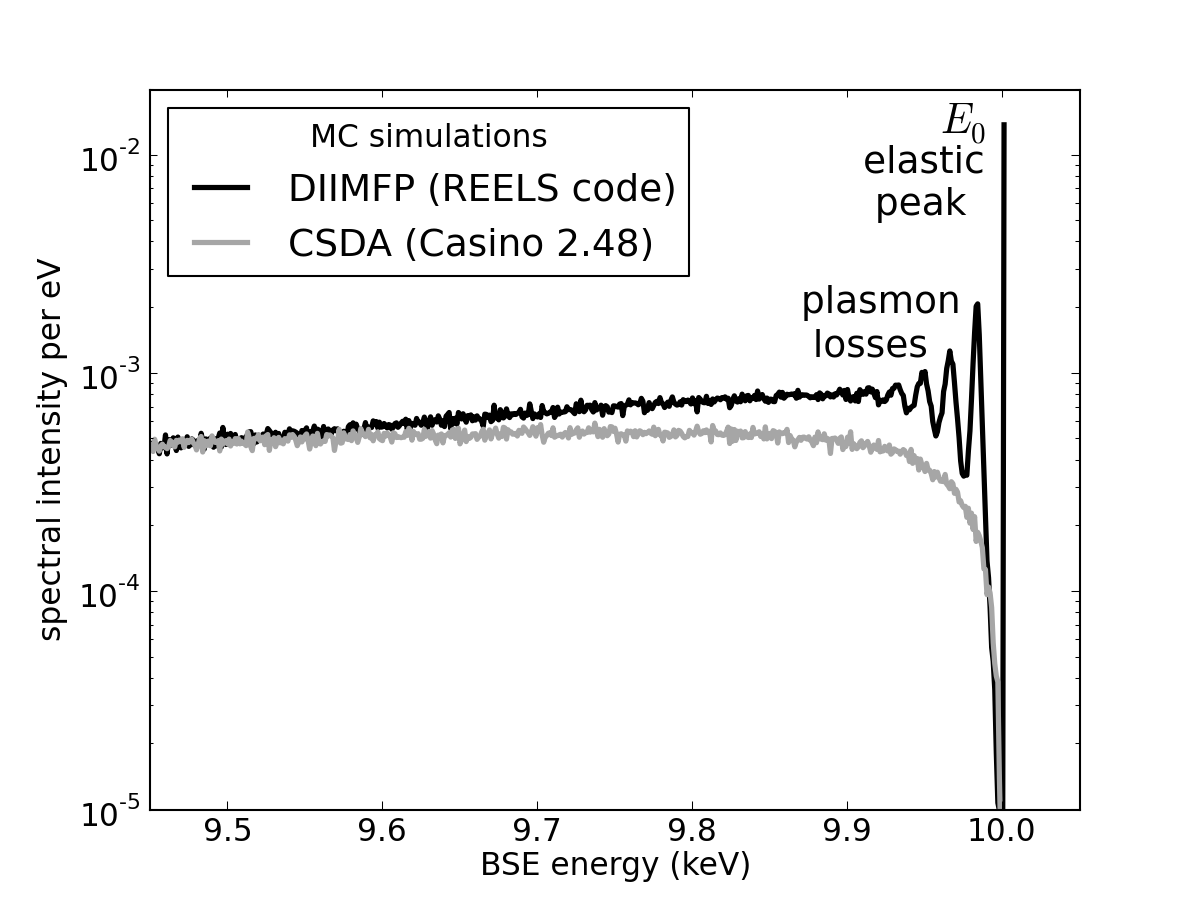}
\caption{Simulated energy spectra of backscattered electrons from Silicon in the continuous slowing down approximation (CSDA) using CASINO \cite{drouin2007scanning} and, alternatively, using the differential inverse inelastic mean free path (DIIMFP) approach \cite{salvatpujol2013sia}. E$_0$=10keV, 70\textdegree{} incidence angle. }
\label{fig:BSEspectrum}
\end{figure}
While the incoherent atomic recoil is the smallest possible energy loss for backscattered electrons, inelastic scattering effects contribute to the formation of Kikuchi patterns in a range of energy losses of up to several hundred eV \cite{deal2008um}. 
This is why a correct description of the energy spectrum of the backscattered electrons is a key ingredient in a more complete model of Kikuchi pattern formation.

Monte-Carlo (MC) simulations provide a useful tool for the description of electron scattering in materials \cite{joy1995mc}, neglecting, however, the influence of diffraction effects.
Inelastic scattering is often taken into account by assigning the loss energy based on the path length travelled in the crystal.
This "continuous slowing down approximation" (CSDA) neglects the quantized character of inelastic loss processes, as a continuous loss energy is always assigned to every path length (no matter how small) and thus the possibility of having no energy loss at all is excluded also (i.e. the coherent elastic backscattering discussed above).

The quantized character of inelastic loss processes can be included by introducing the differential inverse inelastic mean free path (DIIMFP) into MC simulations, which takes into account the specific distribution of possible energy losses for an electron in a material \cite{werner2001sia}. 

\begin{figure}
%\captionsetup{type=figure}% tell subfig package that this is a figure
%\centering
    \subfloat[Angular distribution of BSEs in the hemisphere above the sample, equal-area projection with circles at 10\textdegree{} divisions. The incidence direction at 70\textdegree{} is shown as the small circle. Most intensity is scattered into the forward direction near low takeoff angles.]{
        %\label{subfig:correct}
        \includegraphics[width=7.57cm]{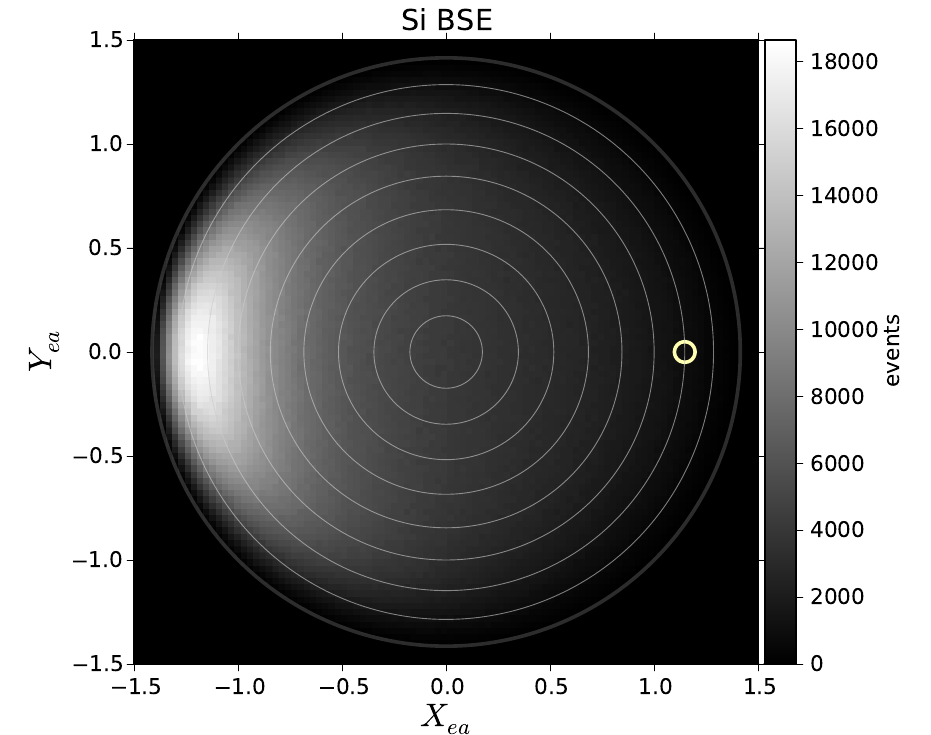}
    } 
%    \subfloat[Intensity at Bragg reflection scaled by $|F_{hkl} |$ ]{
%        %\label{subfig:notwhitelight}
%        \includegraphics[width=7.89cm]{Si_BSE_GN.png}
%    } 
    \hspace{0.5cm}
    \subfloat[Light intensity from a phosphor screen in a typical EBSD setup near the maximum of the angular BSE distribution (gnomonic projection, screen orientation consistent with (a)).]{
        %\label{subfig:notwhitelight}
        \includegraphics[width=6.0cm]{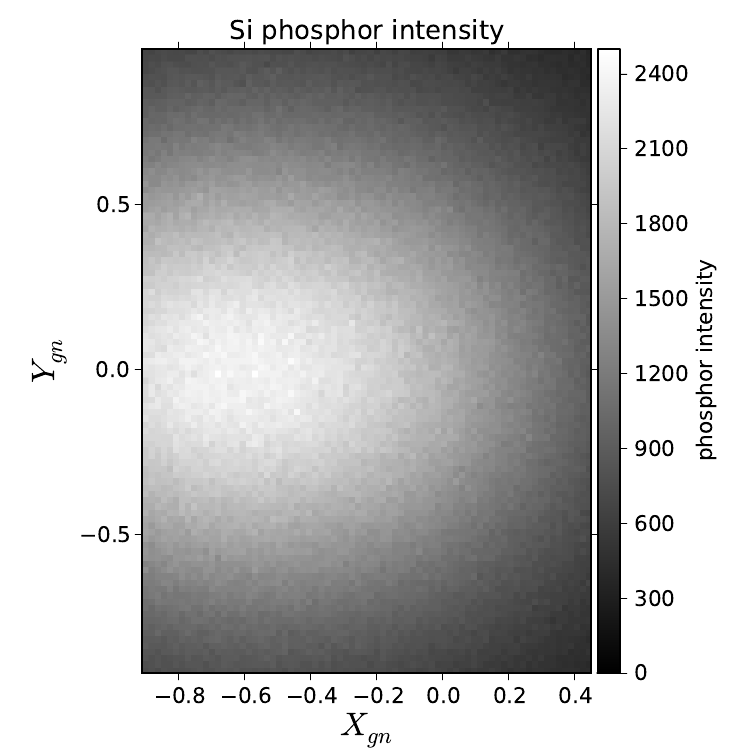}
    }
\caption{Monte-Carlo simulation of electron backscattering from Si. The primary beam energy is 20keV for $50\times10^6$ incident electrons. Backscattered electrons with energy losses up to 10kV are considered.}
\label{fig:BSE}
\end{figure}

In Fig.\,\ref{fig:BSEspectrum}, we calculated the spectrum of electrons backscattered from a Si sample with a primary energy of 10keV at an incidence angle of 70 \textdegree{} \cite{winkelmann2013mm}. The integrated spectral intensity is normalized to 1.0 for the range of 50eV to 10keV. 
In the figure, we show the result of the CSDA approximation as implemented in the CASINO \cite{drouin2007scanning} code compared to a simulation which uses the DIIMFP \cite{salvatpujol2013sia}. 
We can see that the DIIMFP approach reproduces the correct experimental spectrum with an elastic peak and with quantized plasmon losses (17eV for Si) \cite{winkelmann2010njp}. These spectral details are important to exactly model the energy-dependent diffraction effects which are concentrated in the spectral region rather near to the elastic peak. 
This means that the CSDA has strong limitations with respect to the combined problem of electron transport and diffraction, as it neglects the elastic peak, the plasmon features and other energy loss signatures in the backscattered electron energy spectrum and their specific localization in the crystal structure.

The DIIMFP approach can also be used for the simulation of the angular dependence of the backscattered electrons and the intensity distribution on the phosphor screen, as is shown in Fig. \ref{fig:BSE}. For the conversion of electrons into photons, the energy-dependent response function of the phosphor has to be taken into account \cite{ozawa2003cr}. A wealth of information is contained in such angle- and energy-resolved MC simulations, for instance concerning the correlation between scattering angle and the energy-distribution of the backscattered electrons \cite{callahan2013mm} which influences the energy-dependent diffraction pattern structure. 
This type of MC simulation is also relevant to the use of the EBSD screen as an angle-resolved detector of backscattered electrons independently of the actual diffraction effects \cite{payton2013mm,wright2015um,schwarzer2015mt}. 

\begin{figure}   %
\centering
\includegraphics[width=9cm]{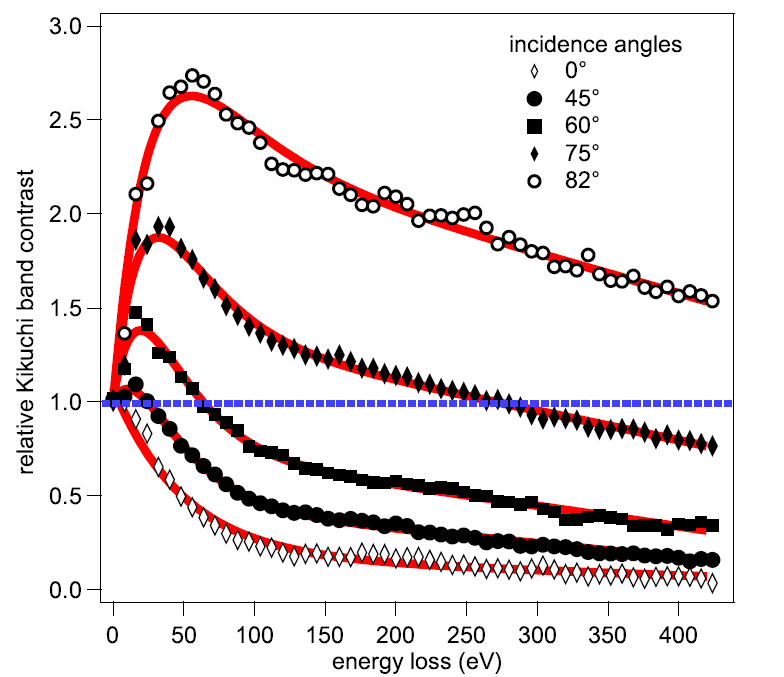}
\caption{Diffraction contrast of backscattered electrons from Si in dependence on the energy loss and the incident beam direction \cite{winkelmann2010njp}. The contrast is measured relative to the quasi-elastically backscattered electrons near zero energy loss (horizontal line, blue).}
\label{fig:kbcontrast}
\end{figure}
The Monte-Carlo simulations are based on the particle picture of electron scattering and thus do not contain diffraction information in a direct way. One of the remaining challenges is to consistently combine the MC model results with the diffraction simulations \cite{callahan2013mm,winkelmann2013mm,winkelmann2010jm}. 
This concerns, for example, the depth sensitivity of EBSD. Based on observations of the width of measured diffraction lines, the energy spread and correspondingly the related depth sensitivity of electrons contributing to an EBSD pattern has been estimated to be in the range between 10nm and 40nm at 20kV, with the lower values reached for dense materials \cite{dingley2004jm}. Experimental observations of the disappearance of Kikuchi pattern diffraction contrast when depositing amorphous layers on crystalline samples are consistent with this estimation \cite{yamamoto77b,zaefferer2007um}. 
A successful model also needs to explain the available experimental data on the energy dependence of Kikuchi band contrast in different scattering geometries \cite{deal2008um,winkelmann2010njp}. 
An example for the complex energy dependence of Kikuchi band contrast from a Si sample is shown in Fig.\,\ref{fig:kbcontrast}, where the data has been measured using a high-resolution electron energy spectrometer \cite{winkelmann2010njp}. This data shows that depending on the incidence angle, Kikuchi patterns can be formed at energy losses up to several hundred eV. 
Due to the concentration of the backscattered intensity near the elastic peak, effectively only a rather narrow energy spectrum contributes to experimental EBSD patterns. 
This justifies the approximation of simulating the dynamical diffraction effects at a fixed "effective" energy near the primary beam energy. For the quantitative reproduction of energy-sensitive effects in the fine structure of experimental EBSD patterns, however, an extended energy range has to be taken into account \cite{winkelmann2009ebsd2}.

Summarizing, it is clearly an important question how the spatial distribution and the energy spectrum of the backscattered electrons in the sample are influencing the EBSD patterns because the inclusion of these effects in dynamical simulations could possibly allow to extract additional information from experimental EBSD measurements. In the the general case, this will require a fully quantum-mechanical treatment of electron backscattering from crystals \cite{dudarev1995prb}.

%\bibliography{library}

\providecommand{\newblock}{}

\end{document}